\documentclass[fleqn,usenatbib]{mnras}

\usepackage{newtxtext,newtxmath}

\newcommand{\Ha}{$\rm{H} \alpha$}
\newcommand{\Hb}{$\rm{H} \beta$}

\newcommand{\PVdblt}{{\rm P}\kern 0.1em{\sc v}~$\lambda\lambda 1117, 1128$}
\newcommand{\CaIIdblt}{{\rm Ca}\kern 0.1em{\sc ii}~$\lambda\lambda 3934, 3969$}
\newcommand{\AlIIIdblt}{{\rm Al}\kern 0.1em{\sc iv}~$\lambda\lambda 1855, 1863$}
\newcommand{\CIVdblt}{{\rm C}\kern 0.1em{\sc iv}~$\lambda\lambda 1548, 1550$}
\newcommand{\MgIIdblt}{{\rm Mg}\kern 0.1em{\sc ii}~$\lambda\lambda 2796, 2803$}
\newcommand{\NVdblt}{{\rm N}\kern 0.1em{\sc v}~$\lambda\lambda 1238, 1242$}  
\newcommand{\SVIdblt}{{\rm S}\kern 0.1em{\sc vi}~$\lambda\lambda 933, 944$} 
\newcommand{\OVIdblt}{{\rm O}\kern 0.1em{\sc vi}~$\lambda\lambda 1031, 1037$} 
\newcommand{\SiIIdblt}{{\rm Si}\kern 0.1em{\sc ii}~$\lambda\lambda 1190, 1193$} 
\newcommand{\SiIVdblt}{{\rm Si}\kern 0.1em{\sc iv}~$\lambda\lambda 1393, 1402$} 
\newcommand{\PV}{\hbox{{\rm P}\kern 0.1em{\sc v}}}
\newcommand{\AlI}{\hbox{{\rm Al}\kern 0.1em{\sc i}}}
\newcommand{\AlII}{\hbox{{\rm Al}\kern 0.1em{\sc ii}}}
\newcommand{\AlIII}{{\hbox{\rm Al}\kern 0.1em{\sc iii}}}
\newcommand{\CaII}{\hbox{{\rm Ca}\kern 0.1em{\sc ii}}}
\newcommand{\CII}{\hbox{{\rm C}\kern 0.1em{\sc ii}}}
\newcommand{\CIIe}{\hbox{{\rm C$^{\ast}$}\kern 0.1em{\sc ii}}}
\newcommand{\CIII}{\hbox{{\rm C}\kern 0.1em{\sc iii}}}
\newcommand{\CIV}{\hbox{{\rm C}\kern 0.1em{\sc iv}}}
\newcommand{\CV}{\hbox{{\rm C}\kern 0.1em{\sc v}}}
\newcommand{\HI}{\hbox{{\rm H}\kern 0.1em{\sc i}}}
\newcommand{\HII}{\hbox{{\rm H}\kern 0.1em{\sc ii}}}
\newcommand{\Lya}{\hbox{{\rm Ly}\kern 0.1em$\alpha$}}
\newcommand{\Lyb}{\hbox{{\rm Ly}\kern 0.1em$\beta$}}
\newcommand{\Lyg}{\hbox{{\rm Ly}\kern 0.1em$\gamma$}}
\newcommand{\Lyd}{\hbox{{\rm Ly}\kern 0.1em$\delta$}}
\newcommand{\Lye}{\hbox{{\rm Ly}\kern 0.1em$\epsilon$}}
\newcommand{\Lyphi}{\hbox{{\rm Ly}\kern 0.1em$\phi$}}
\newcommand{\Lyfive}{\hbox{{\rm Ly}\kern 0.1em$5$}}
\newcommand{\Lysix}{\hbox{{\rm Ly}\kern 0.1em$6$}}
\newcommand{\Lyseven}{\hbox{{\rm Ly}\kern 0.1em$7$}}
\newcommand{\Lyeight}{\hbox{{\rm Ly}\kern 0.1em$8$}}
\newcommand{\Lynine}{\hbox{{\rm Ly}\kern 0.1em$9$}}
\newcommand{\Lyten}{\hbox{{\rm Ly}\kern 0.1em$10$}}
\newcommand{\Lyeleven}{\hbox{{\rm Ly}\kern 0.1em$11$}}
\newcommand{\HeI}{\hbox{{\rm He}\kern 0.1em{\sc i}}}
\newcommand{\HeII}{\hbox{{\rm He}\kern 0.1em{\sc ii}}}
\newcommand{\FeI}{\hbox{{\rm Fe}\kern 0.1em{\sc i}}}
\newcommand{\FeII}{\hbox{{\rm Fe}\kern 0.1em{\sc ii}}}
\newcommand{\FeIII}{\hbox{{\rm Fe}\kern 0.1em{\sc iii}}}
\newcommand{\MnII}{\hbox{{\rm Mn}\kern 0.1em{\sc ii}}}
\newcommand{\MgI}{\hbox{{\rm Mg}\kern 0.1em{\sc i}}}
\newcommand{\MgIb}{\hbox{{\rm Mg}\kern 0.1em{\sc i}}\kern 0.05em{\rm b}}
\newcommand{\MgII}{\hbox{{\rm Mg}\kern 0.1em{\sc ii}}}
\newcommand{\MgIII}{\hbox{{\rm Mg}\kern 0.1em{\sc iii}}}
\newcommand{\NI}{\hbox{{\rm N}\kern 0.1em{\sc i}}}
\newcommand{\NII}{\hbox{{\rm N}\kern 0.1em{\sc ii}}}
\newcommand{\NIII}{\hbox{{\rm N}\kern 0.1em{\sc iii}}}
\newcommand{\NV}{\hbox{{\rm N}\kern 0.1em{\sc v}}}
\newcommand{\OVI}{\hbox{{\rm O}\kern 0.1em{\sc vi}}}
\newcommand{\OI}{\hbox{{\rm O}\kern 0.1em{\sc i}}}
\newcommand{\OII}{\hbox{[{\rm O}\kern 0.1em{\sc ii}]}}
\newcommand{\OIII}{\hbox{[{\rm O}\kern 0.1em{\sc iii}]}}
\newcommand{\OIV}{\hbox{{\rm O}\kern 0.1em{\sc iv}]}}
\newcommand{\SI}{{\rm S}\kern 0.1em{\sc i}}
\newcommand{\SIV}{{\rm S}\kern 0.1em{\sc iv}}
\newcommand{\SVI}{{\rm S}\kern 0.1em{\sc vi}}
\newcommand{\SiI}{\hbox{{\rm Si}\kern 0.1em{\sc i}}}
\newcommand{\SiII}{\hbox{{\rm Si}\kern 0.1em{\sc ii}}}
\newcommand{\SiIII}{\hbox{{\rm Si}\kern 0.1em{\sc iii}}}
\newcommand{\SiIV}{\hbox{{\rm Si}\kern 0.1em{\sc iv}}}
\newcommand{\SII}{\hbox{{\rm S}\kern 0.1em{\sc ii}}}
\newcommand{\SIII}{\hbox{{\rm S}\kern 0.1em{\sc iii}}}
\newcommand{\NaI}{\hbox{{\rm Na}\kern 0.1em{\sc i}}}
\newcommand{\NaID}{\hbox{{\rm Na}\kern 0.1em{\sc i}}\kern 0.05em{\rm D}}
\newcommand{\TiII}{\hbox{{\rm Ti}\kern 0.1em{\sc ii}}}
\newcommand{\kms}{\hbox{~km~s$^{-1}$}}
\newcommand{\cmsq}{\hbox{cm$^{-2}$}}

\usepackage[T1]{fontenc}

\DeclareRobustCommand{\VAN}[3]{#2}
\let\VANthebibliography\thebibliography
\def\thebibliography{\DeclareRobustCommand{\VAN}[3]{##3}\VANthebibliography}

\usepackage{graphicx}	
\usepackage{amsmath}	

\title[Galaxy quenching caused by a lack of CGM]{Evidence for galaxy quenching in the green valley caused by a lack of a circumgalactic medium}

\author[G. G. Kacprzak et al.]{Glenn G. Kacprzak$^{1,2}$\thanks{E-mail: gkacprzak@swin.edu.au}, 
Nikole M. Nielsen$^{1,2}$, 
Hasti Nateghi$^{1,2}$, 
Christopher W. Churchill$^{3}$,
\newauthor Stephanie K. Pointon$^{2}$,
Themiya Nanayakkara$^{1}$, 
Sowgat Muzahid$^{4}$, 
Jane C. Charlton$^{5}$\\
$^{1}$Centre for Astrophysics and Supercomputing, Swinburne University of Technology, Hawthorn, Victoria 3122, Australia\\
$^{2}$ARC Centre of Excellence for All Sky Astrophysics in 3  Dimensions (ASTRO 3D)\\
$^{3}$Department of Astronomy, New Mexico State University, Las Cruces, NM 88003, USA\\
$^{4}$Leibniz-Institute for Astrophysics Potsdam (AIP), An der Sternwarte 16, D-14482 Potsdam, Germany\\
$^{5}$Department of Astronomy and Astrophysics, The Pennsylvania State University, State College, PA 16801, USA\\
}

\date{Accepted XXX. Received YYY; in original form ZZZ}

\pubyear{2020}

\begin{document}
\label{firstpage}
\pagerange{\pageref{firstpage}--\pageref{lastpage}}
\maketitle

\begin{abstract}
The relationship between a galaxy's properties and its circumgalactic medium (CGM) provides a unique view of how galaxies evolve. We present an interesting edge-on ($i=86$~degrees) disk galaxy (G1547) where the CGM is probed by a background quasar at a distance of 84 kpc and within 10 degrees of the galaxy major axis.  G1547 does not have any detectable CGM absorption down to stringent limits, covering {\HI} (EW$_r$<0.02~{\AA}, log(N({\HI})$/$cm$^{-2}$)<12.6) and a range of low and high ionisation absorption lines ({\OI}, {\CII}, {\NII}, {\SiII}, {\CIII}, {\NIII}, {\SiIII}, {\CIV}, {\SiIV}, {\NV} and {\OVI}). This system is rare, given the covering fraction of $1.00_{-0.04}^{+0.00}$ for sub-L* galaxies within 50-100~kpc of quasar sightlines. G1547 has a low SFR (1.1~M$_{\odot}$~yr$^{-1}$), SSFR ($1.5\times10^{-10}$~yr$^{-1}$)  and $\Sigma_{\rm SFR}$ ($0.06$~M$_{\odot}$~yr$^{-1}$~kpc$^{-2}$) and does not exhibit AGN or star-formation driven outflows. Compared to the general population of galaxies, G1547 is in the green valley and has an above average metallicity with a negative gradient. When compared to other {\HI} absorption-selected galaxies, we find that quiescent galaxies with log(SSFR$/$yr$^{-1}$)$<-11$ have a low  probability (4/12) of possessing detectable {\HI} in their CGM, while all  galaxies (40/40) with log(SSFR$/$yr$^{-1}$)$>-11$ have {\HI} absorption. We conclude that SSFR is a good indicator of the presence of {\HI} CGM. Interestingly however, G1547 is the only galaxy with log(SSFR$/$yr$^{-1}$)$>-11$ that has no detectable CGM. Given the properties of G1547, and its absent CGM, it is plausible that G1547 is undergoing quenching due to a lack of accreting fuel for star-formation, with an estimated quenching timescale of $4\pm1$~Gyr. G1547 provides a unique perspective into the external mechanisms that could explain the migration of galaxies into the green valley.
\end{abstract}



\section{Introduction}

The multi-phase circumgalactic medium (CGM) is ubiquitous among all types of galaxies. It has been attributed to arise within outflows, accretion, tidal streams, and in group and cluster environments \citep[e.g.,][]{lopez08,ho17,peroux16,chen1127,schroetter19,hamanowicz20}. These physical processes occur within the CGM and are assumed to be responsible for shaping the stellar mass-metallicity relation \citep[e.g.,][]{tremonti04,mannucci10,kacprzak16}, the stellar mass to halo mass relation \citep[e.g.,][]{behroozi10}, and the halo mass-star formation rate relation \citep[e.g.,][]{behroozi13}. Therefore, understanding the processes ongoing within the CGM provides direct insight into galaxy formation and evolution.   

Background quasars have been used to detect a range of ions from the CGM, such as {\HI}, {\MgII}, {\CIV}, {\OVI}, which have been used to ascertain its origins \citep{chen01a,lundgren09,chen10a,rudie12,magiicat1,werk13,borthakur15,pointon19}. It has been widely demonstrated that the presence of absorption can be quite telling as to where the CGM comes from. However, the lack of CGM absorption also provides useful hints into understanding galaxies.

 It has been shown that galaxies with low specific star formation rates of log(SSFR$/$yr$^{-1}$)$<-11$ have little or no {\OVI} absorption \citep{tumlinson11}. The lack of {\OVI} could be caused by either a lack of outflows due to an absence of active star-formation \citep{tumlinson11}, or caused by a larger group environment \citep{pointon17}, or could be due to galaxies residing in larger halo masses that further ionise oxygen to higher undetectable levels \citep{oppenheimer16}.   This also is consistent with the lack of {\CIV} absorption found in other galaxy surveys. \citet{burchett16} reports that both low mass galaxies (log(M/M$_{\odot})<9.5)$ and galaxies in larger group environments have less than  a 10\% probability of possessing  log(N({\CIV})$/$\cmsq)$>13.5$ absorption within the CGM. Therefore the lack of absorption can be quite telling as to the physical processes or environments that are responsible for removing or ionising the CGM around galaxies, which has a direct effect on their evolution. 
 
 Nonetheless, {\HI} detected through {\Lya} absorption seems to be ever present around galaxies \citep{werk13,borthakur15,burchett16, pointon19}. The covering fraction is extremely high and extends out to impact parameters of $>300$~kpc, which encompasses everything from the ISM to the CGM and IGM. \citet{rudie12} has shown that absorption systems with N({\HI})$>10^{14.5}$~cm$^{-2}$ are tightly correlated with galaxies while lower N({\HI}) is correlated with galaxies only on Mpc scales. This indicates that low N({\HI}) is likely from the IGM. Thus, given the vast extent of {\HI} absorption, there are very few known non-absorbing systems from typical galaxies reported in the literature. In one such case, \citet{johnson14} reported a transparent sightline associated with a pair of interacting mature galaxies (log(SSFR$/$yr$^{-1}$)$\sim$ $-11.2$ and $-11.7$) within 20~kpc of a quasar sightline at $z=0.12$. The sightline had a log(N({\HI})$/$\cmsq) limit of $<13.65$ and no detected metal-lines. It was suggested that the absence of absorption so close to the quasar sightline was caused by gas reservoirs being significantly depleted or removed by galaxy-galaxy interactions. This unique system highlights the potential impact of galaxy interactions on the CGM. Therefore, any case in which a galaxy exhibits a lack of {\HI} absorption would be an interesting indicator of an unusual state of the CGM, which could reveal insights on how these galaxies evolve.
 
Here, we present a unique galaxy that is nearly perfectly edge-on with its major axis aimed almost directly at the quasar sightline at $D=84$~kpc. This orientation has generally been demonstrated to have a higher rate of detection of low and high ionisation absorption \citep{bouche12,kcn12,kacprzak15}. However, this galaxy (G1547) does not have any metal or {\HI} absorption to very sensitive limits (e.g., log(N(\HI)$/$\cmsq)$<12.6$). We explore the properties of G1547 and compare them to the distribution of properties of a general sample of galaxies to investigate what this implies for the evolution of galaxies. The paper is organised as follows: In \S~\ref{sec:data} we describe the data and analysis. In \S~\ref{sec:results} we present the results and absorption and galaxy properties. In \S~\ref{sec:discussion}, we discuss the implications of the lack of absorption and conclude that G1547 is likely undergoing quenching. We present our concluding remarks in \S~\ref{sec:conclusion}. Throughout we adopt an H$_{\rm 0}=70$~\kms Mpc$^{-1}$, $\Omega_{\rm M}=0.3$, $\Omega_{\Lambda}=0.7$ cosmology.

\section{Observations and Analysis}
\label{sec:data}
We present the data and analysis of the SDSS~J154743.53+205216.6 quasar field (J1547). We focus this paper on an edge-on galaxy at $z=0.13536$, named G1547, that has its major axis aligned with the quasar sightline at $D=84$~kpc and has no associated CGM absorption. The image of the field is presented in Fig.~\ref{fig:image} and the galaxy properties are presented in Table~\ref{tab:props}.

\subsection{Quasar HST/COS Spectra}

{\it HST}/COS quasar spectra were obtained using the G130M (exposure time of 9841 seconds), G160M (10,254 seconds)  and G185M (2609 seconds) as part of the ``Multiphase Galaxy Halos'' Survey [from PID 13398 \citep{kacprzak15,muzahid15,muzahid16,nielsen17,pointon17,kacprzak19a,kacprzak19b,ng19,pointon19,pointon20}]. These spectra cover a large range of metal and hydrogen absorption lines at the redshift of G1547. The spectra were reduced using the {\sc CALCOS} software. Individual grating integrations were aligned and co-added using the code `coadd\_x1d' created by \citet{danforth10}\footnote{\url{http://casa.colorado.edu/~danforth/science/cos/costools.html}}. We binned the spectra by three pixels to increase the signal-to-noise and our analysis was performed on the binned spectra. The continuum normalisation was obtained by fitting the absorption-free regions with low-order polynomials.

We searched for absorption lines within $\pm$1000~{\kms} of the galaxy redshift. The COS spectrum covered a range of possible ions, such as {\HI}, {\OI} {\CII}, {\NII}, {\SiII}, {\CIII}, {\NIII}, {\SiIII}, {\CIV}, {\SiIV}, {\NV} and {\OVI}, yet we did not detect any significant absorption. We calculated 3$\sigma$ upper limits on the equivalent widths associated with the non-absorbing G1547 galaxy, which are presented in Table~\ref{tab:abs}. To compute the column density limits presented in Table~\ref{tab:abs}, we assumed a single cloud with a Doppler parameter of $b=8$~{\kms} for the lower ionisation states \citep[adopted from][]{pointon19} and 30~{\kms} for {\NV} and {\OVI} \citep[adopted from][]{pointon17}. We note however that the column density limits derived here are insensitive to the choice of Doppler parameter in this regime of the linear part of the curve of growth. 

\subsection{Galaxy Imaging and Models}

A {\it HST}/WFPC2 image was taken of this field using the F606W filter with an exposure time of 1800 seconds (PID:5099). The image was reduced using the WFPC2 Associations Science Products Pipeline \citep[WASPP, see][]{kacprzak11b}.  {\it HST} photometry, quoted in the AB system, was computed using the Source Extractor software \citep[SExtractor;][]{bertin96} with a detection criterion of 1.5$\sigma$ above background. 

Galaxy morphological properties were modelled with a two-component disk$+$bulge using GIM2D \citep{simard02}, where the disk has an exponential profile while the bulge has a S{\'e}rsic profile with $0.2\leq n\leq 4.0$. We apply the standard convention of an azimuthal angle $\Phi=0^{\circ}$ defined to be along the galaxy projected major axis and $\Phi=90^{\circ}$ defined to be along the galaxy projected minor axis. The modelled galaxy morphological parameters are presented in Table\ref{tab:props}.

Pan-STARRS \citep{chambers16} $grizy$ imaging and photometry were also available, having 5$\sigma$ depths of 23.3, 23.2, 23.1, 22.3, 21.3 in each filter, respectively.  The galaxy photometry was measured on the catalogue images in the same manner as the {\it HST} image. These magnitudes are presented in Table~\ref{tab:props}.

\subsection{Galaxy Spectroscopy}

A spectroscopic survey of the field was performed near the quasar using the Keck Echelle Spectrograph and Imager, ESI \citep{sheinis02}.  We binned the CCD by two in the spatial direction, which yields pixel scales of $0.27-0.34''$ over the echelle orders of interest. Also, we binned the CCD by two in the spectral direction resulting in a resolution of 22~\kms~pixel$^{-1}$ (${\rm FWHM}\sim90$~km/s) for a $1''$ slit.  ESI has a wavelength coverage of 4000--10,000~{\AA}, which allows for the  detection of a large range of emission and absorption lines for galaxies in this field such as the {\OII} doublet, $\rm{H}\beta$, {\OIII} doublet, {\MgIb}, {\NaID}, $\rm{H}\alpha$, and the [\NII] doublet.

The ESI data were reduced using IRAF. Galaxy spectra were both vacuum and heliocentric velocity corrected to provide a direct comparison with the quasar spectra.  The derived wavelength solution was verified against a catalogue of known sky-lines which resulted in a RMS difference of $\sim0.03$~{\AA} ($\sim2$~{\kms}).  
The initial flux calibration was performed using a flux calibration star observed on the same night. We then applied spectrophotometric calibration using the {\it HST} F606W photometry to ensure absolute flux calibration, within 10\%, and to account for any light loss from the slit \citep[see][for details]{nanayakkara16}. 

The Gaussian fitting software FITTER \citep[see][]{archiveI} was used to determine galaxy emission and absorption line fluxes and equivalent widths. Emission-line fluxes are presented in Table~\ref{tab:props} along with additional quantities derived using them, which are discussed in the next Section. 

To measure the stellar mass we used the spectral energy distribution modelling code, FAST \citep{kriek09}. We modelled both the galaxy spectrum and Pan-STARRS photometry together. We applied a Galactic extinction correction to the photometry using a two-dimensional dust reddening map \citep{plank11,planckdust14} and assuming a \cite{cardelli89} attenuation law. We have used stellar population synthesis models from \cite{bc03} computed with a \cite{chabrier03} initial mass function and assumed a delayed exponentially declining star formation history. We also fixed the stellar extinction to $A_V=1$, which were computed from G1547's spectrum (see Table~\ref{tab:props}). The stellar mass of G1547 is log(M$_*/$M$_{\odot}$)$=9.7\pm0.1$.

\begin{figure*}
	\includegraphics[width=\textwidth]{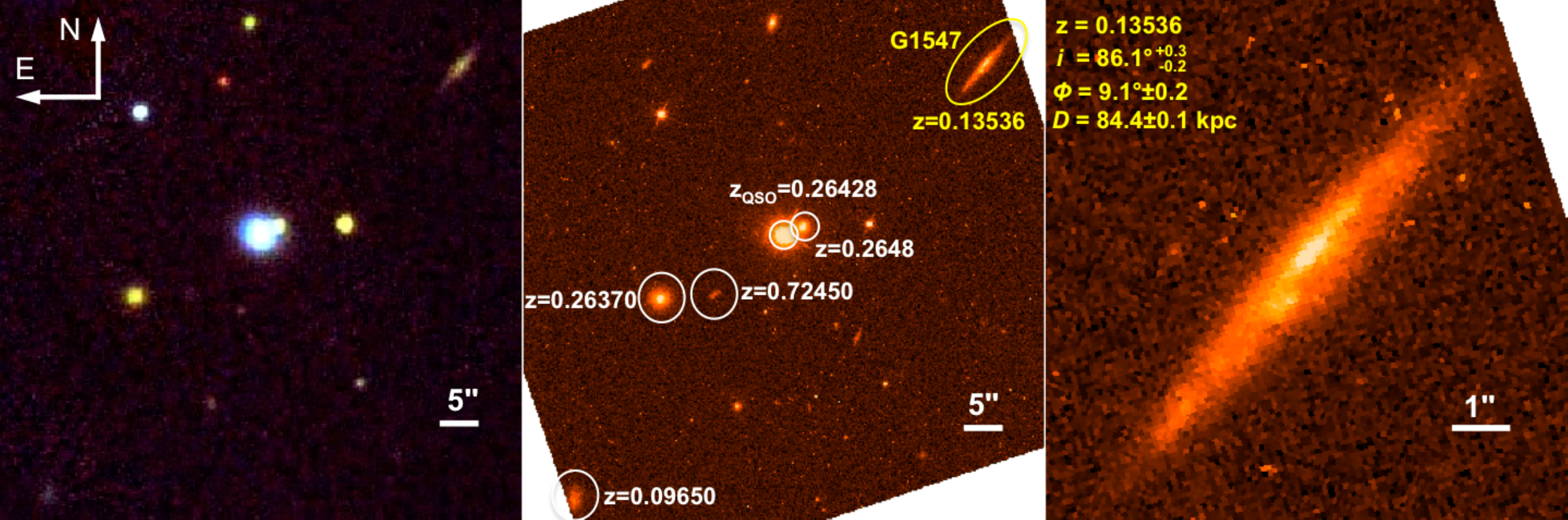}
    \caption{(left:) Pan-STARRS 70$\times$70$''$ $gri$ colour image of the field with the quasar in the centre and the foreground G1547 edge-on disk galaxy in the upper right. (Centre:) {\it HST}/WFPC-2 F606W image of the same 70$\times$70$''$ field is shown along with the spectroscopic redshifts that are labelled for each object. Note G1547 at $z=$0.13536 has its major axis almost directly aligned with the quasar sightline ($\Phi=9.1$ degrees). (Right:) A {\it HST} 10$\times$10$''$ image zoom of the nearly edge-on ($i=86.1$ degrees) bulgeless disk galaxy with the GIM2D-modelled orientation properties listed in the upper-left. }
    \label{fig:image}
\end{figure*}
\begin{figure*}
	\includegraphics[width=17cm]{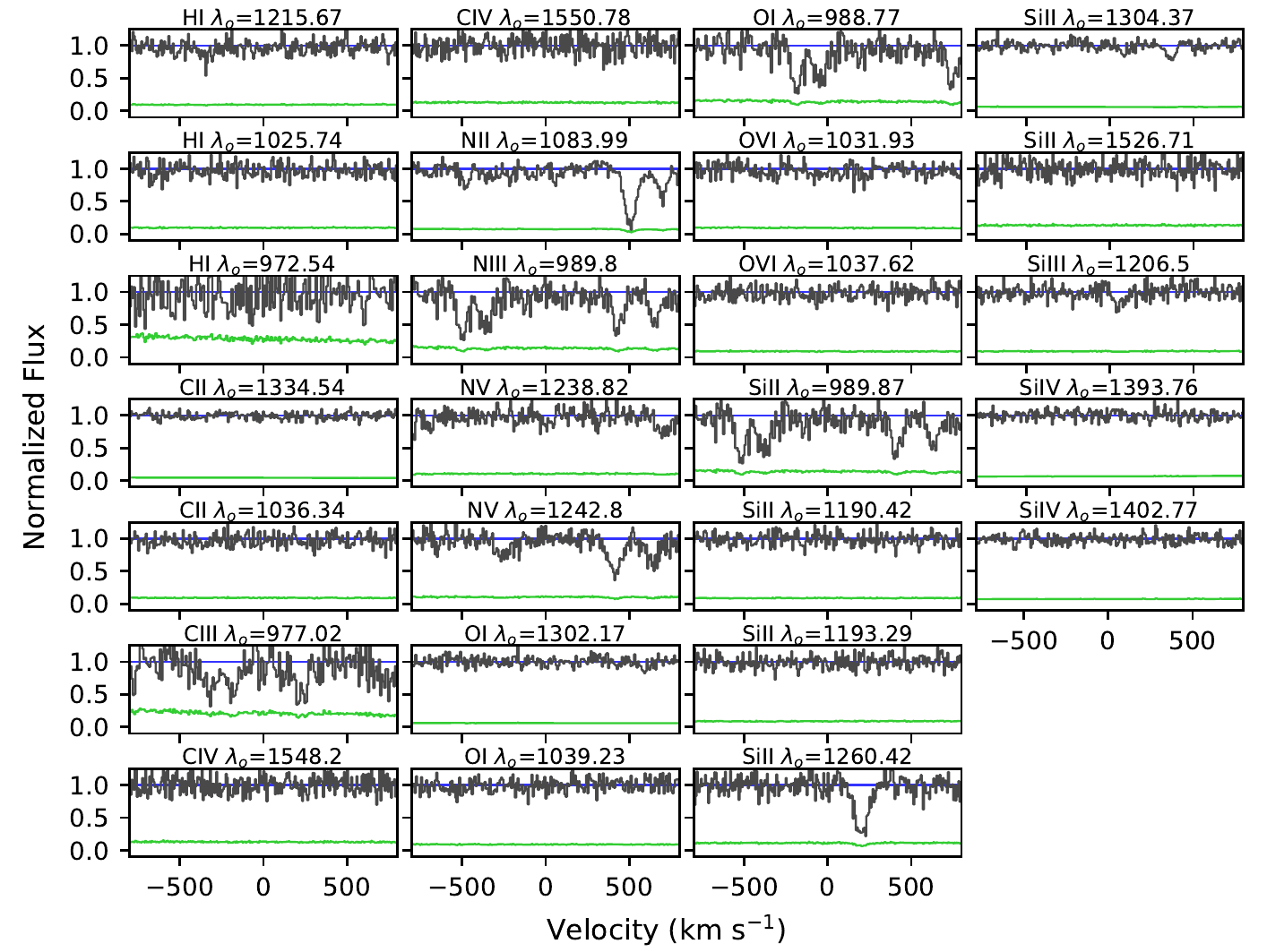}
    \caption{{\it HST}/COS spectrum shown in black for each ion, which is labelled above each panel. The green line shows the sigma spectrum and the blue line is the fit to the continuum. The velocity zero-point is set to the systemic redshift of G1547. Note that we do not detect any {\HI} absorption (log(N(\HI)$/$\cmsq)$<$12.6) and we do not detect any metal-lines at any ionisation state over a large velocity range. Rest-frame equivalent widths and column density limits are listed in Table~\ref{tab:abs}.}
  \label{fig:abs}
\end{figure*}
\begin{table}
	\centering
	\caption{Properties of the edge-on disk galaxy G1547.}
	\label{tab:props}
	\begin{tabular}{llc} 
		\hline
		Properties& Notation (Units) & Value\\
		\hline
		redshift              & $z$     &0.13536$\pm$0.00002 \\
	    stellar mass          & log(M$_*/$M$_{\odot})$ & 9.7$\pm$0.1 \\
		impact parameter      & $D$ (kpc)&  84.4$\pm$0.1 \\
		disk inclination      & $i$ (degrees)& 86.1$^{+0.3}_{-0.2}$   \\
		bulge-to-total        & $B/T$ & 0.00$^{+0.01}_{-0.00}$   \\
		azimuthal angle       & $\Phi$ (degrees)& 9.1$\pm$0.2\\
	    half light radius     & $r_h$ (kpc) & 5.8$\pm$0.1 \\
	    {\it HST} magnitude         & $m_{F606W}$ & 19.46$\pm$0.03\\
	    Pan-STARRS magnitudes & $m_g$ & 21.07$\pm$0.13 \\
	                          & $m_r$ & 20.02$\pm$0.04 \\
	                          & $m_i$ & 19.34$\pm$0.02 \\
	                          & $m_z$ & 19.17$\pm$0.04 \\
	                          & $m_y$ & 19.13$\pm$0.06 \\
	    line fluxes $(\times10^{-17})$  & F(\Ha)~ (erg~cm$^{-2}$~s$^{-1}$~\AA$^{-1}$)  & 40.1$\pm$2.6\\
	                          & F(\NII~$\lambda$6585)~  & 16.2$\pm$3.5\\
                              & F(\OIII~$\lambda$5008)~    &3.9$\pm$1.5 \\
	                          & F(\Hb)~ & 8.0$\pm$0.7\\
	   colour excess: {\HII} regions             & $E(B-V)_{\rm HII}$ & 0.6$\pm$0.1 \\
	   extinction: {\HII} regions                & $A_V,_{\rm HII}$ & 1.5$\pm$0.3 \\
	   colour excess: stellar            & $E(B-V)_{*}$ & 0.26$\pm$0.04 \\
	   extinction: stellar          & $A_V,_{*}$ & 1.0$\pm$0.2 \\

	    star formation rate   & SFR (M$_{\odot}$~yr$^{-1}$)& 1.1$\pm$0.3 \\
	    specific SFR          & SSFR (yr$^{-1}$)& 1.5$\pm$0.4$\times$$10^{-10}$ \\
	    SFR surface density& $\Sigma_{\rm SFR}$ (M$_{\odot}$~yr$^{-1}$~kpc$^{-2}$) & 0.06$\pm$0.02 \\
	    ISM metallicity PP04 & log(O/H)+12 & 8.68$\pm$0.05  \\
	    \hline
	\end{tabular}
\end{table}
\begin{table}
	\centering
	\caption{CGM Absorption rest-frame equivalent width and column density limits for the G1547 $z=0.13536$ edge-on disk galaxy. We use the highest oscillator strength transitions to compute the column density limits.}
	\label{tab:abs}
	\begin{tabular}{lrcc} 
		\hline
		Ion & Wavelength &  EW$_r$ (3~$\sigma$) & log(N$(X)$/$\cmsq)$\\
		   &   (\AA) &  (\AA) & \\
		\hline
\HI  &972.54    & $<$0.041  &   \\
\HI &1025.74    & $<$0.018  &   \\
\HI &1215.67    & $<$0.019  &   $<$12.6\\
\CII &1036.34   & $<$0.017  &   \\
\CII &1334.54   & $<$0.010  &   $<$12.7\\
\CIII &977.02   & $<$0.038  &   $<$12.9\\
\CIV &1548.20   & $<$0.030  &  $<$12.9\\
\CIV &1550.78   & $<$0.033  &   \\
\NII &1083.99   & $<$0.014  &   $<$13.2\\
\NIII &989.80   & $<$0.025  &   $<$13.5\\
\NV &1238.82    & $<$0.022  &   $<$13.1\\
\NV &1242.80    & $<$0.021  &   \\
\OI & 1302.17   & $<$0.014  &  $<$13.3\\
\OI & 1039.23   & $<$0.016  &   \\
\OI & 988.77    & $<$0.030  &   \\
\OVI &1031.93   & $<$0.018  & $<$13.1  \\
\OVI &1037.62   & $<$0.017  &   \\
\SiII &989.87   & $<$0.023  &   \\
\SiII &1190.42  & $<$0.018  &   \\
\SiII &1193.29  & $<$0.018  &   \\
\SiII &1260.42  & $<$0.023  &  $<$12.3 \\
\SiII &1304.37  & $<$0.013  &   \\
\SiII &1526.71  & $<$0.033  &   \\
\SiIII &1206.50 & $<$0.019  & $<$12.3  \\
\SiIV &1393.76  & $<$0.015  & $<$12.3  \\
\SiIV &1402.77  & $<$0.016  &   \\
	    \hline
	\end{tabular}
\end{table}

\section{Results}
\label{sec:results}
In Fig.~\ref{fig:image}, we show a 70$\times$70$''$ Pan-STARRS $gri$ colour image and a {\it HST} F606W image of the field. The quasar ($z=0.26428$) is located in the centre of the image and near two galaxies at similar redshifts placing them in a group. We have identified a background galaxy at $z=0.72450$ and two foreground galaxies at $z=0.09650$ and $z=0.13536$ (G1547). 

The $z=0.09650$ galaxy has been discussed in \citet{chen01a} and in \citet{pointon19} and has an impact parameter of 79.8~kpc. This is a blue ($B-K=1.0$) edge-on disk galaxy ($i=80.9$ degrees) that has its major axis $\Phi=54.7$ degrees offset from the quasar sightline. It has no detectable metal-line absorption but has Ly$\alpha$ absorption with log(N(\HI)$/$\cmsq)=13.75$\pm0.03$ \citep{pointon19}. 

The G1547 $z=0.13536$ galaxy was first identified by \citet{chen01a}. Here we have obtained a more accurate redshift and have now measured a wide range of galaxy properties that will be the focus of this paper. This galaxy is almost perfectly edge-on with an inclination angle of $i=86.1^{+0.3}_{-0.2}$ degrees with no observable bulge component, having a budge-to-total ratio of $B/T=0.00^{+0.01}_{-0.00}$. The galaxy resides at an impact parameter $84.4\pm0.1$~kpc and has its major axis nearly aligned with the quasar sightline with $\Phi=9.1\pm0.2$ degrees. 

From a quasar absorption line perspective, having an edge-on disk galaxy with its major axis aligned with the quasar sightline provides the ideal geometry to probe different physical processes such as gas accretion, which has been shown to co-rotate and fall along the major axis of galaxies \citep[see review from][]{kacprzak17}. Also given that there are known inclination dependencies on the CGM properties \citep[e.g.,][]{bordoloi11,kacprzak11b,magiicat5}, edge-on disks are ideal case studies.

Fig.~\ref{fig:abs} shows the large range of ions covered by the {\it HST}/COS spectra for G1547. However, there are no detectable {\HI} or metal-lines to quite sensitive limits. The rest-frame 3$\sigma$ equivalent width upper limits of the metal lines range from 0.01 to 0.04~{\AA}, and the Lyman-$\alpha$ upper limit is 0.019~{\AA}. All of the rest-frame equivalent upper limits are presented in Table~\ref{tab:abs}. The Lyman-$\alpha$ equivalent width upper limit of 0.019~{\AA} translates to a log(N(\HI)$/$\cmsq)$\leq$12.6. All column density upper limits are also presented in Table~\ref{tab:abs}. The lack of absorption seems surprising given the galaxy's moderate impact parameter and geometry, where absorption has been shown to preferentially to be found \citep{bouche12,kcn12,kacprzak15}.  


\begin{figure*}
	\includegraphics[width=\textwidth]{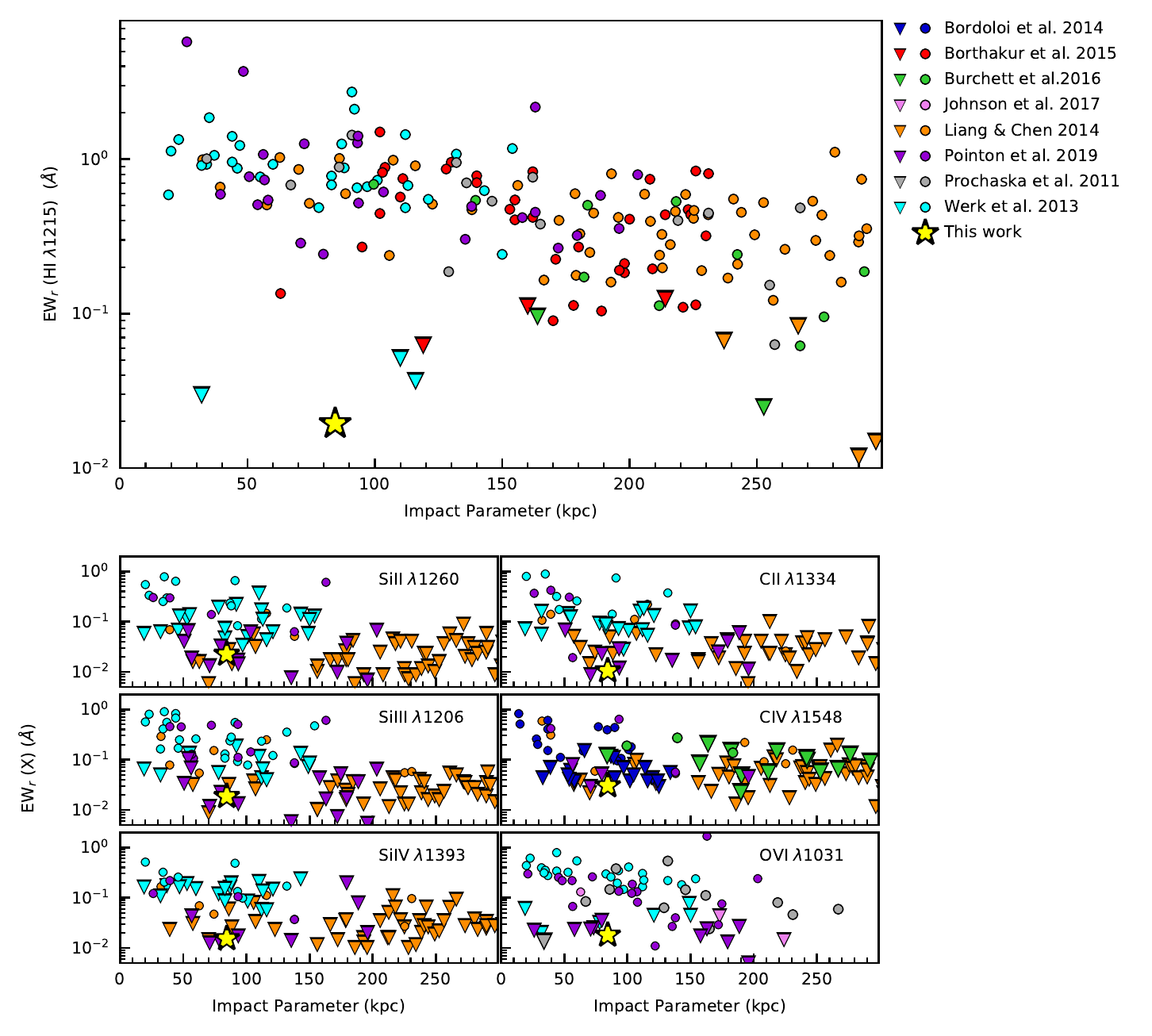}
    \caption{Rest-frame equivalent widths (circles) and 3$\sigma$ upper-limits (triangles) shown as a function of impact parameter for a range of ions as listed. These systems are selected from literature \citep{prochaska11,werk13,bordoloi-cosdwarfs,liang14,borthakur15,burchett16,johnson17,pointon19} and show only galaxies in the stellar mass range of $9\leq$log(M$_*/$M$_{\odot}$)$\leq$11 (or luminosity range of $0.1\leq$log(L$_*/$L$_{\odot}$)$\leq$1 since masses are not available in \citealt{prochaska11}). The yellow stars indicate equivalent width upper limits for G1547. Note that G1547 is one of only two {\HI} non-absorbers within 100~kpc.}
    \label{fig:profiles}
\end{figure*}

\begin{figure*}
	\includegraphics[width=\textwidth]{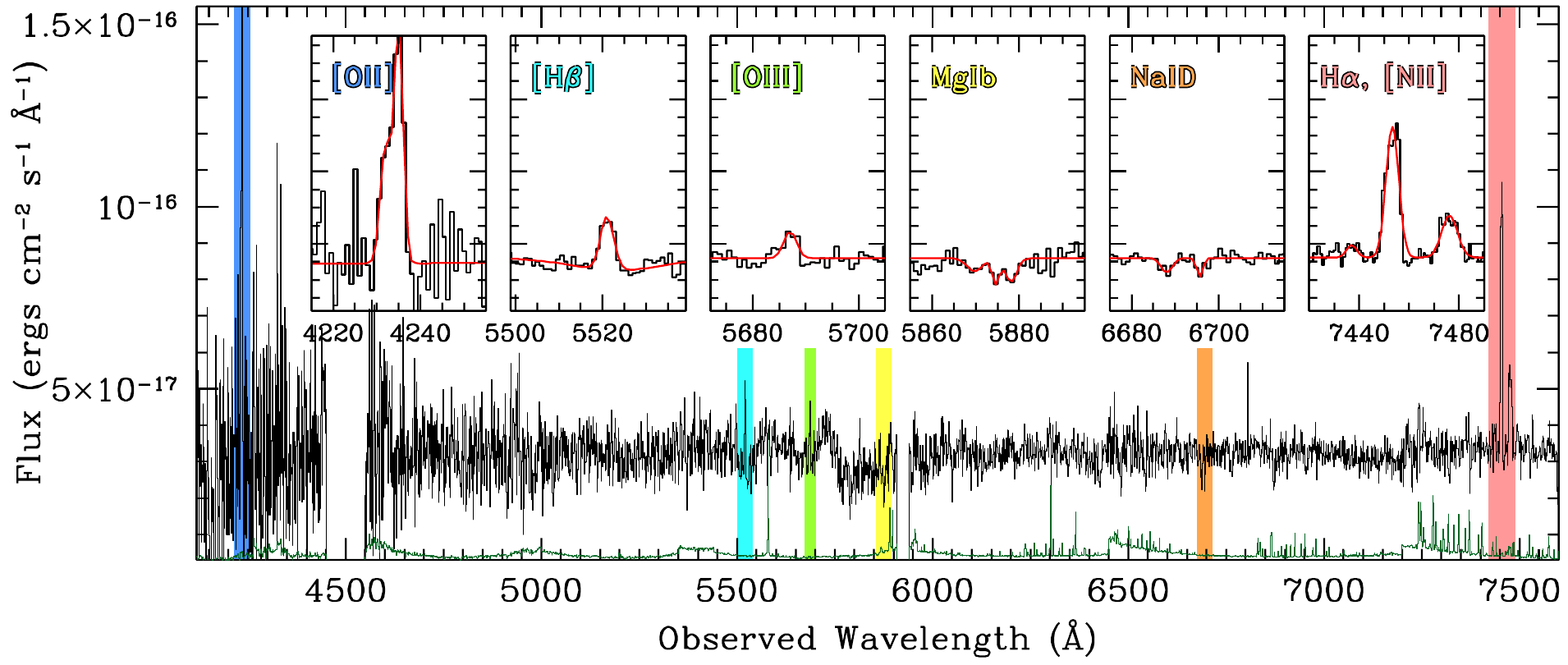}
    \caption{A subset of observed wavelengths from the flux calibrated Keck/ESI spectrum of the $z=0.13536$ edge-on disk galaxy G1547 is shown. The data shown here are binned in the spectral direction by three pixels. The spectrum (black) and the uncertainty spectrum (green) are shown along with highlighted emission and absorption-line regions of interest that are colour-coded to the six zoom-in panels. Note the full range of emission and absorption lines used to determine the galaxy's metallicity, SFR and outflow properties.}
    \label{fig:galspec}
\end{figure*}

\begin{figure*}
	\includegraphics[width=\textwidth]{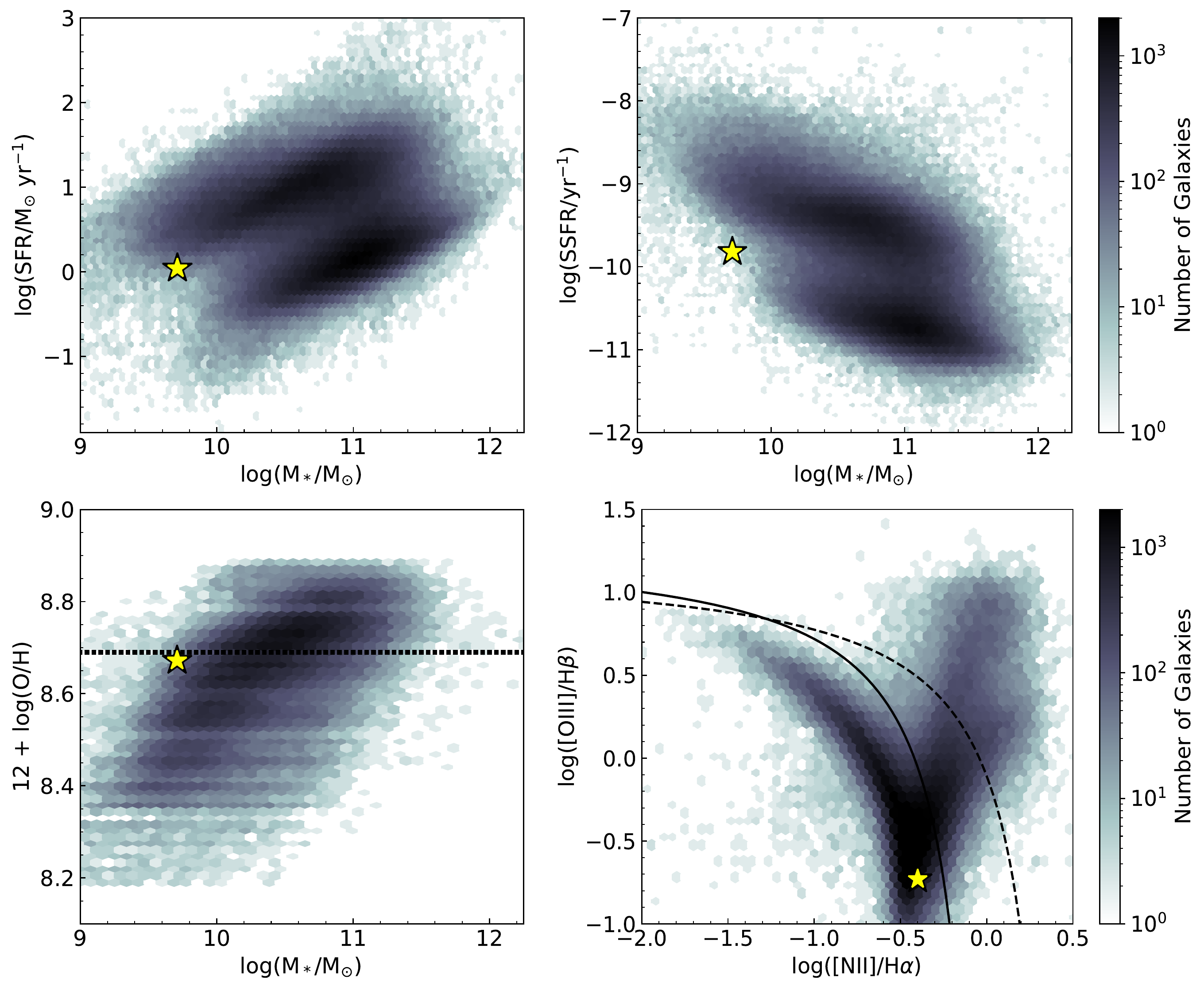}
    \caption{Properties shown for over 490,000 galaxies with  $0.05\leq z_{\rm spec} \leq 0.15$, which are taken from the MPA-JHU catalogue. The yellow star indicates the measured properties of G1547. (Upper-left): SFR as a function of stellar mass. Both blue and red sequences can be seen, with our galaxy residing at the low mass and low SFR region. (Upper-right:) SSFR as a function of stellar mass. G1547 resides at a SSFR between the blue and red sequence, within the green valley. (Bottom-left:) The stellar mass and ISM metallicity relation. Using \citet{kewley08}, we normalise the metallicities from \citet{tremonti04} to H$\alpha$, NII metallicity indicator from \citet{pettini04} in order to directly compare to G1547. The dotted horizontal line is the solar abundance \citep{asplund09}. Note the above average metallicity for G1547 at fixed mass. (Bottom-right:) The BPT diagram for the SDSS galaxies that have {\OIII}, {\Hb}, {\NII}, and {\Ha} with $SNR\geq 5$. The solid curve shows the \citet{kauffmann03} empirical division between star-forming galaxies and AGNs and the dashed curve shows the \citet{kewley01} theoretical division. G1547 is a typical galaxy with no current AGN activity.}
    \label{fig:properties}
\end{figure*}

To demonstrate the uniqueness of the lack of absorption associated with this galaxy, we show the {\HI} and metal-line radial profiles for absorbing and non-absorbing galaxies out to impact parameters of 300~kpc in Fig.~\ref{fig:profiles}. Galaxy surveys from the literature containing CGM measurements, limits and galaxy masses are plotted for comparison \citep{prochaska11,werk13,liang14,bordoloi-cosdwarfs,borthakur15,burchett16,johnson17,pointon19}. Given there exists a mass dependence on metal-line column densities and profiles \citep[e.g.,][]{chen10a,cwc-masses,magiicat3,tumlinson13,oppenheimer16,ng19}, we only show galaxies between a stellar mass range of $9\leq$log(M$_*/$M$_{\odot}$)$\leq$11 (or luminosity range of $0.1\leq$log(L$_*/$L$_{\odot}$)$\leq$1 since masses are not available in \citealt{prochaska11}). We do not limit the redshift range as there is no known evolution of the radial profiles \citep[e.g.,][]{chen12} and also the studies shown here primarily contain galaxies with $z\lesssim0.3$. All rest-frame equivalent width limits were adjusted such that they are quoted at 3$\sigma$.

We show the {\Lya} equivalent width as a function of impact parameter in the top panel of Fig.~\ref{fig:profiles}. There exists a well-defined anti-correlation seen here, with a high gas covering fraction out to 300~kpc. There are 13 galaxies with upper-limits, with 7 of those having stringent limits below 0.05{\AA} out of a total of 258 galaxies. Regardless, there are very few non-absorbing galaxies over a large range of impact parameters. 

Most notably, there are only two galaxies, out of 72 galaxies that have no detectable {\HI} down to a limit of 0.05~{\AA} within 100~kpc. Within  0$-$50~kpc, we compute a covering fraction of  $0.96_{-0.09}^{+0.04}$. The covering fraction 1$\sigma$ errors are derived from binomial statistics \citep{gehrels86}. Excluding our galaxy, the covering fraction in the 50$-$100~kpc bin is $1.00_{-0.04}^{+0.00}$. Thus, our galaxy has a 4\% chance likelihood of existing within the 1$\sigma$ errors for that given covering fraction. Including the galaxy, we determine an actual covering fraction of $0.98_{-0.05}^{+0.02}$ for galaxies with $9\leq$log(M$_*/$M$_{\odot}$)$\leq$11 at 50$-$100~kpc. 
The lowest impact parameter non-absorbing galaxy was identified in \citet{werk12} as "bonus" galaxy 1342$-$0053\_77\_10 at $z=0.2013$. It is a red galaxy ($u-r>$2.45) with no measurable star-formation (SFR$<0.08$~M$_{\odot}$~yr$^{-1}$), having a stellar mass of log(M$_*$/M$_{\odot})=$10.50 and resides at an impact parameter of 32~kpc \citep{werk12}. This galaxy has no detectable {\HI} or metal-lines. \citet{kacprzak15} modelled the {\it HST} image of this galaxy, which appears to be an oblate S0-type galaxy (see their figure~2) with an inclination angle of 72~degrees and an azimuthal angle of 45 degrees. Given the type of galaxy and the location of the quasar sightline with respect to the galaxy, it might not be unexpected that this galaxy is lacking absorption \citep{bordoloi11,bouche12,kcn12}. We will expand on why this low impact parameter galaxy from \citet{werk12}, along with other galaxies, may not exhibit CGM absorption later in Section~\ref{sec:discussion} (also see Fig~\ref{fig:nonabs}).

G1547 is the only non-absorbing galaxy within the 50$-$100~kpc range with a very stringent rest-frame equivalent width upper limit of 0.019~{\AA}. As previously mentioned, it is curious as to why G1547 does not have any detectable {\HI} given its disk geometry and orientation with respect to the quasar sightline. There are three additional galaxies that do not host {\HI} absorption between 100--150~kpc, which we will also discuss the Section~\ref{sec:discussion}.

In Fig.~\ref{fig:profiles}, we also show a selection of low and high ionisation metal-lines for absorbing and non-absorbing galaxies. Note that the majority of the metal-lines in the low ionisation states, i.e., {\SiII}, {\SiIII}, {\CIII} and {\SiIV}, are found within 150~kpc, while higher ionisation states, i.e., {\CIV} and {\OVI}, are found out to nearly 300~kpc. G1547 has some of the best rest-frame equivalent width upper limits, however it is not unique in its lack of metals at these impact parameters. The only ion where upper limits are less common at low impact parameter is {\OVI} as there are only 9 non-absorbing galaxies within 100~kpc. It has been noted that a lack of {\OVI} could either be attributed to the lack of star formation \cite{tumlinson11}, or due to a halo mass and {\OVI} column density dependence \citep{ng19,oppenheimer16}. We next explore additional measurable properties of G1547 with the goal of inferring why it lacks significant absorption.

Fig.~\ref{fig:galspec} shows a Keck/ESI spectrum of G1547. This galaxy has a large range of emission and absorption-lines indicating that it is a star-forming galaxy. Galaxy properties computed using its spectrum can be found in Table~\ref{tab:props} and a subset of those properties are shown in Fig.~\ref{fig:properties}.
Although dust corrections for low-mass star-forming galaxies like G1547 do not depend on inclination \citep[e.g.,][]{wang18}, there are likely systematics affecting the derivation of some galaxy properties such as the ISM metallicity and SFR \citep[see][]{devour17}. An absolute correction is not possible here without additional deep infrared data that is not available for this field. However, our conclusions would likely remain unchanged even if some of the galaxy properties were to change as predicted in \citet{devour17}.

We derived an attenuation-corrected {\Ha} star formation rate using relations from \citet{tran15} for dust corrections.  We computed and applied attenuation corrections for both the nebular emission \citep{cardelli89} and stellar continuum \citep{calzetti00}. Combining the {\Ha} and {\Hb} flux ratios, and using the \citet{cardelli89} attenuation curve for the diffuse interstellar medium, where $R_V=3.1$, $k($\Ha$)=2.53$ and $k($\Hb$)=3.61$, we computed a $E(B-V)_{\rm HII}=0.6\pm0.1$ and $A_{V,\rm HII}=1.5\pm0.3$. We then used the observed stellar-to-nebular attenuation ratio of $E(B-V)_{*} = 0.44E(B-V)_{\rm HII}$ \citep{calzetti00} and a R$_V=4.05$ to compute a  $E(B-V)_{*}=0.26\pm0.04$ and $A_{V,*}=1.0\pm0.2$ for the stellar attenuation. Applying both those attenuation corrections to the {\Ha} luminosity, a corrected star formation rate of SFR=$1.1\pm0.3$~M$_{\odot}$~yr$^{-1}$ was obtained using log(SFR) = log($L$(\Ha))$-41.27$ from \citet{hao11}.

In Fig.~\ref{fig:properties} (upper-left), the SFR as a function of stellar mass is shown, with G1547 indicated by a yellow star. The grey density points show the properties of a comparison sample of $\sim$490,000 galaxies selected from the MPA-JHU catalogue having spectroscopic redshifts of $0.05\leq z \leq 0.15$. The MPA-JHU catalogue provides SFRs, SSFRs, stellar masses, metallicities and emission line fluxes based on the methods of  \citet{brinchmann04}, \citet{kauffmann03} and \citet{tremonti04}. Given G1547's low SFR of 1.1$\pm$0.3~M$_{\odot}$~yr$^{-1}$, it resides above the red sequence and below the star-formation sequence within the green valley.

In Fig.~\ref{fig:properties} (upper-right), the specific star formation rate as a function of stellar mass is shown. Again G1547 resides between both the red and blue sequences with a SSFR$=1.5\pm0.4\times10^{-10}$~yr$^{-1}$. Note that \citet{tumlinson11} showed that nearly all (26/29) star-forming galaxies exhibit {\OVI} CGM absorption while the majority of quiescent galaxies (8/12, SSFR$<1\times10^{-11}$ yr$^{-1}$) do not exhibit {\OVI} CGM absorption. Thus, it is interesting that this particular galaxy does not have any detectable {\OVI} absorption down to a column density limit of log(N(\OVI)$/$\cmsq)$<13.1$ (which is lower than any limit in \citealp{tumlinson11}) yet has a SSFR above $1\times10^{-11}$ yr$^{-1}$. We will discuss this further in Section~\ref{sec:discussion}.



In Fig.~\ref{fig:properties} (lower-left), the ISM metallicity as a function of the stellar mass is shown. We compute a gas-phase oxygen abundance for each galaxy using the N2 relation of \citet{pettini04}, where 12+log(O/H)=8.90+0.57$\times$log(\NII/{\Ha}).  The  MPA-JHU catalogue provides metallicities derived using the methods from \citet{tremonti04}, which estimates the metallicity statistically, based on the fluxes of all prominent emission lines combined with population synthesis and photo-ionisation modelling. In order to directly compare to our computed value of the galaxy metallicity, we apply the metallicity conversions from \citet{kewley08}. These conversions allow metallicities that have been derived using different calibrations to be converted to the same base calibration, which results in agreement with the M-Z relation to within 0.03~dex on average.

\begin{figure}
	\includegraphics[width=\columnwidth]{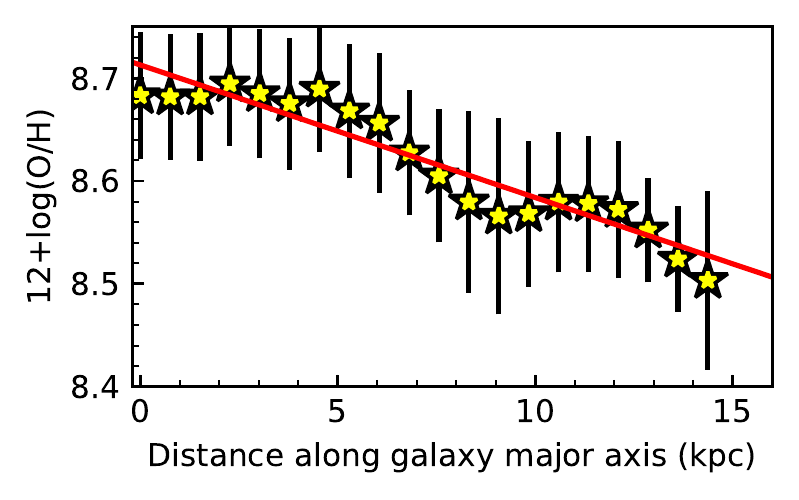}
    \caption{The major axis averaged metallicity profile for G1547. A fit to the gradient results in 12$+$log(O$/$H)$=-0.013\pm0.001\times R+8.71\pm0.01$, where $R$ is the distance along the major axis of the galaxy in kpc. G1547 exhibits a shallow gradient that is consistent with most star-forming and S0 galaxies \citep[see review][]{sanchez20}.}
    \label{fig:metalprofile}
\end{figure}

G1547 has a metallicity of 12+log(O/H)=8.68$\pm$0.05 using the N2 method. We also derived the metallicity using the O3N2 method \citep{pettini04} and found a consistent result of $8.70\pm0.06$. We find that G1547 resides above the expected metallicity for its stellar mass by $\sim$0.25~dex and is consistent with a solar metallicity of 8.69. The higher-than-typical metallicity likely indicates that the galaxy is not undergoing any significant low metallicity gas accretion at this stage and its ISM metallicity is increasing from the ongoing star-formation \citep[e.g.,][]{kacprzak16}. It has been noted that flat metallicity gradients could be a signature of gas inflows/accretion \citep{kewley10,jones13}. In Fig.~\ref{fig:metalprofile}, we show the major axis metallicity profile for G1547. A fit to the data yields 12$+$log(O$/$H)$=-0.013\pm0.001\times R+8.71\pm0.01$, where $R$ is the distance along the major axis of the galaxy in kpc. G1547 exhibits a shallow gradient that is consistent with most star-forming and S0 galaxies \citep[see review][]{sanchez20}. In general, \citet{sanchez20} showed that there is a large scatter in the slopes of galaxy ISM metallicity gradients. Thus it is difficult to draw conclusions regarding G1547 even though it exhibits a negative gradient. However, it is plausible that G1547 has not undergone any recent accretion.  

Fig.~\ref{fig:properties} (lower-right), shows the BPT diagram for the SDSS galaxies that have {\OIII}, {\Hb}, {\NII}, and {\Ha} detected with SNR$\geq$5. The solid curve shows the \citet{kauffmann03} empirical division between star-forming galaxies and AGNs. The dashed curve shows the \citet{kewley01} theoretical division. AGNs reside above these lines while star-forming galaxies reside below these lines. G1547 resides at log(\NII/\Ha)$=-0.40\pm0.1$ and log(\OIII/\Hb)$=-0.7\pm0.2$, placing it well within the range expected for typical galaxies with no significant AGN activity at present and thus no active AGN-driven outflows.

\begin{figure}
	\includegraphics[width=\columnwidth]{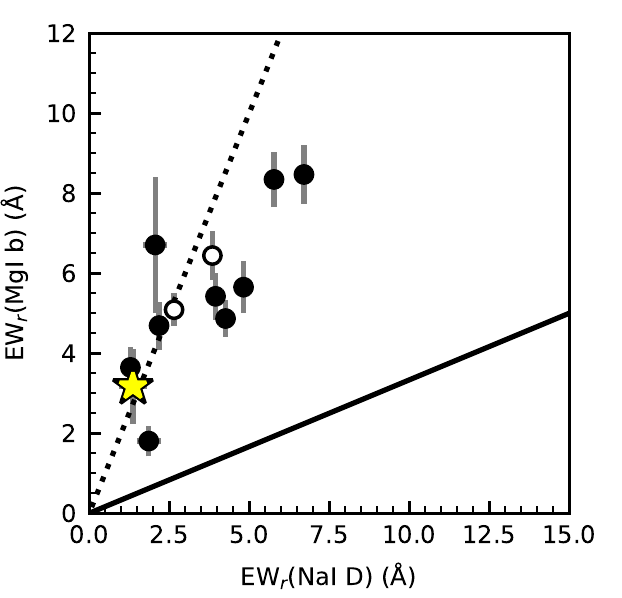}
    \caption{Rest-frame equivalent widths of the {\NaID} (stellar$+$ISM) and {\MgIb} (stellar) absorption lines. Below the solid line 80\% of the galaxies have outflows while only 25\% of galaxies above the solid line have outflows \citep{rupke05a,rupke05b}. The dashed line shows the expected stellar contribution to the {\NaID} absorption by scaling the absorption strength of {\MgIb}. The black points are $z\sim0.1$ {\MgII} absorbers (full circle) and non-absorbing (open circle) galaxies from \citep{kacprzak11kin}. The yellow star is G1547, which resides on top of the dashed line, indicating stellar contributions only to the {\NaID} absorption and no outflows are expected to be present.}
    \label{fig:outflows}
\end{figure}
We now explore the possibility that G1547 could have an expected CGM reservoir populated by star formation driven outflows. We investigate the possible evidence of outflows using three methods.  We first compute the galaxy star-formation surface density and compare to expected thresholds for star formation driven outflows. We then compute {\MgIb} (stellar) and {\NaID} (stellar+ISM) line ratios in order to quantify the possibility of outflows. We then explore possibility of velocity offsets between {\NaID} absorption and the {\Ha} emission line as an additional test for the presence of outflows.

G1547 is forming stars at a rate of 1.1$\pm$0.3~M$_\odot$~yr$^{-1}$. Combined with the galaxy half light radius of 5.8$\pm$0.1~kpc, with an ellipticity of 0.83, yields a star formation surface density of $\Sigma_{\rm SFR}=$0.06$\pm$0.02~M$_{\odot}$~yr$^{-1}$~kpc$^{-2}$. It has been demonstrated that outflows appear to be ubiquitous in galaxies where the SFR per unit area exceeds $\Sigma_{\rm SFR}=0.1$~M$_{\odot}$~yr$^{-1}$~kpc$^{-2}$ \citep{heckman03,heckman11,sharma17}. G1547 is a factor of 1.7 short of this threshold, suggesting that it is unlikely G1547 is currently generating stellar driven outflows.

Another good tracer of outflows is obtained from the {\NaID} and {\MgIb} absorption lines.  Due to their similar ionisation potentials (5.14~eV for {\NaID}, 7.65~eV for {\MgIb}), both peak in absorption strength in the spectra of cool K–M stars \citep{jacoby84}. While {\MgIb} is purely stellar in origin, the {\NaID} resonance line can also be absorbed by the ISM and is commonly detected as entrained gas within galactic winds \citep{heckman00,martin05,rupke05a,rupke05b}. Therefore, the ratio of these two lines has been used to successfully separate outflowing galaxies from galaxies with little-to-no winds \citep[e.g.,][]{heckman00}.

In Fig.~\ref{fig:outflows}, we show the {\NaID} and {\MgIb} rest-frame equivalent widths for G1547 along with a sample of $z\sim0.1$ {\MgII} absorbing (full circles) and non-absorbing galaxies (open circles) from \citet{kacprzak11kin}. The dashed line is the expected stellar contribution of {\MgIb} as function of {\NaID}, where  $EW_r($\NaID$) = 0.75\times EW_r($\MgIb$)$ \citep{heckman00}. The solid line gives the approximate location of starburst galaxies with strong outflows observed by \citet{rupke05a,rupke05b}. It was found that 80\% of galaxies below this line ($EW_r($\NaID$) = 3\times EW_r($\MgIb$)$) contained strong outflows, whereas only 25\% of galaxies found above this line contained outflowing gas.  G1547 resides directly on the relation where {\NaID} is expected to have a purely stellar origin with little-to-no contribution from entrained outflowing gas. 

Another way to detect outflows is to measure velocity offsets between galaxy emission and absorption lines \citep[e.g.,][]{heckman00,rupke05b,weiner09,steidel10,rubin10}. Here we compare the velocity difference between {\Ha} and {\NaID} since \citet{heckman00} determined that galaxies with {\NaID} absorption found within 70~{\kms} of the systematic velocity were consistent with a predominantly stellar origin. Furthermore, galaxies with {\NaID} absorption found  blueshifted greater than 100~{\kms} were associated with outflows $\sim$70\% of the time.
For G1547, we find a mean {\NaID} velocity offset of $\Delta v =-32\pm19$~{\kms}, which is consistent with a predominantly stellar origin even though the line is mildly blueshifted with respect to the galaxy.  

It is worth noting that, since G1547 is viewed with an edge-on orientation, and even if outflows were present, we have a reduced ability for detecting them. It has been noted in the literature that outflows are more easily identified in galaxies with low inclination angles \citep[e.g.,][]{bordoloi14,heckman00,korni12,rubin-winds14}. Thus it could be possible that we are unable to directly detect or rule out star-formation driven outflows. However, the SFR and $\Sigma_{\rm SFR}$ are well below what is expected for star-formation driven outflows \citep{bordoloi14} and it remains reasonable to infer that G1547 does not contain significant and far reaching outflows.

\begin{figure*}
	\includegraphics[width=\textwidth]{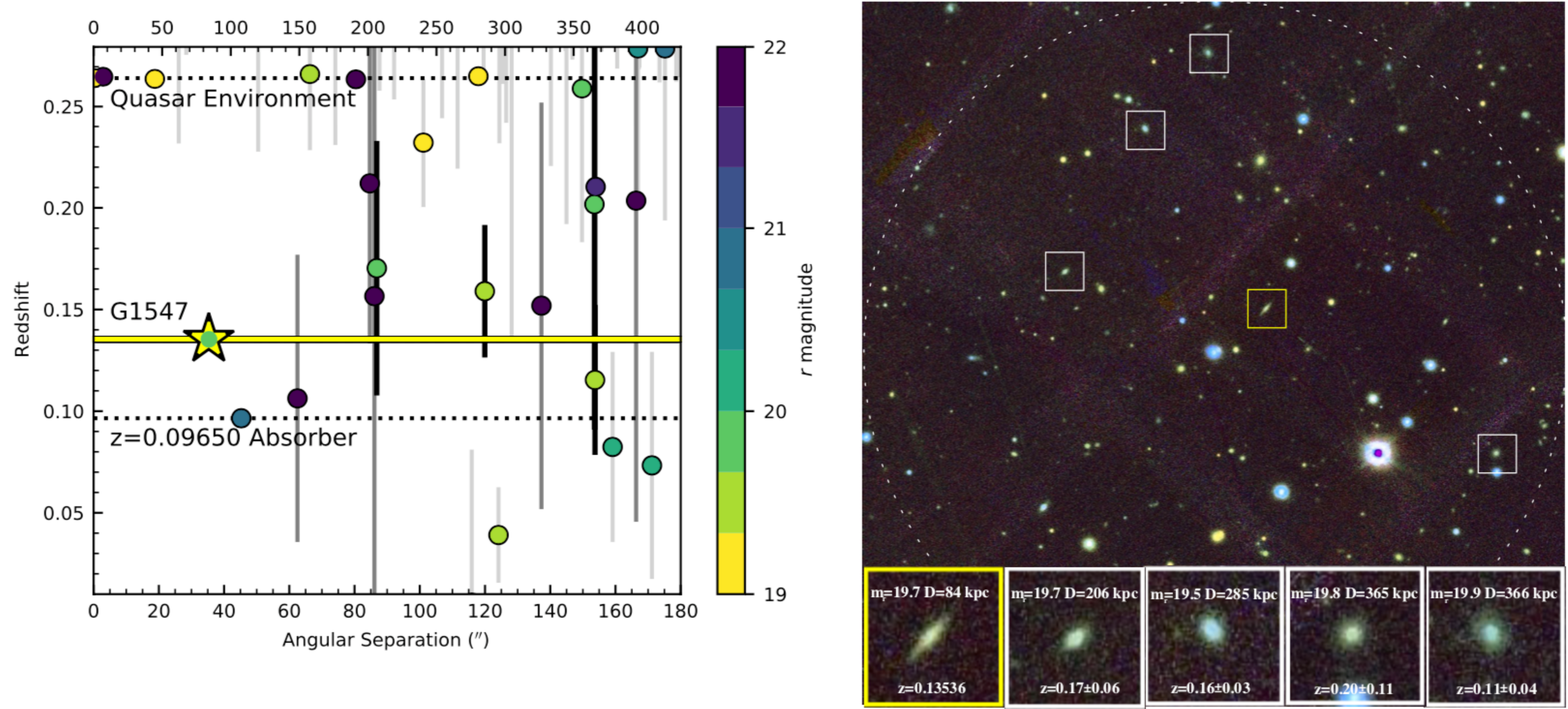}
    \caption{(left) Spectroscopic and photometric redshifts within 3$'$ in radius from the quasar. The data are colour-coded as function of galaxy apparent $r-$magnitude. The G1547 galaxy is highlighted (yellow line) along with the quasar redshift (dashed line, where an over-density of galaxies can be seen) and the $z=0.09650$ absorbing galaxy. Galaxies with photometric redshifts consistent with G1547 have dark grey (faint galaxies with $m_r\geq20$) and black error bars ($m_r<20$). Both the angular scale and the impact parameter scale are shown, with the latter assuming the redshift of G1547 (2.28~kpc/$''$). Note that at this redshift a $\delta z = 0.05$ is equivalent to a $\delta v = 13,000$~{\kms} or co-moving radial distance of 200~Mpc. Furthermore, the angular diameter distance scale changes significantly at these redshifts such that at $z=0.1$ the scale is 0.98~kpc/$''$ and at $z=0.25$ the scale is 3.91$''$~kpc/$''$. Thus, the physical separation should be taken as a guide and not a true spatial distribution of the galaxies. Although the error bars are large, there are 5 dwarf galaxies ($m_r\geq21.5$) that could potentially be at a similar redshift. There are also 4 galaxies of similar brightness as G1547 that have large error bars. Also note that 4/5 dwarfs and 1/4 brighter galaxies are also consistent with the $z=0.09650$ absorption system.  (Right) $6\times6'$ Pan-STARRS $gri$ image of the field, with a dotted circle indicating a radius of 3$'$ around the quasar located in the centre of the image. The four other possible bright galaxies are highlighted with their photometric redshifts, $m_r$, and impact parameters (assuming the redshift of G1547) labelled. Although there could be a larger-scale association, no other massive galaxy dominates this system and all other potential members are at least 2.5 times further from the quasar than G1547. Thus we treat this system as an isolated galaxy.}
    \label{fig:env}
\end{figure*}

\begin{figure*}
	\includegraphics[width=\textwidth]{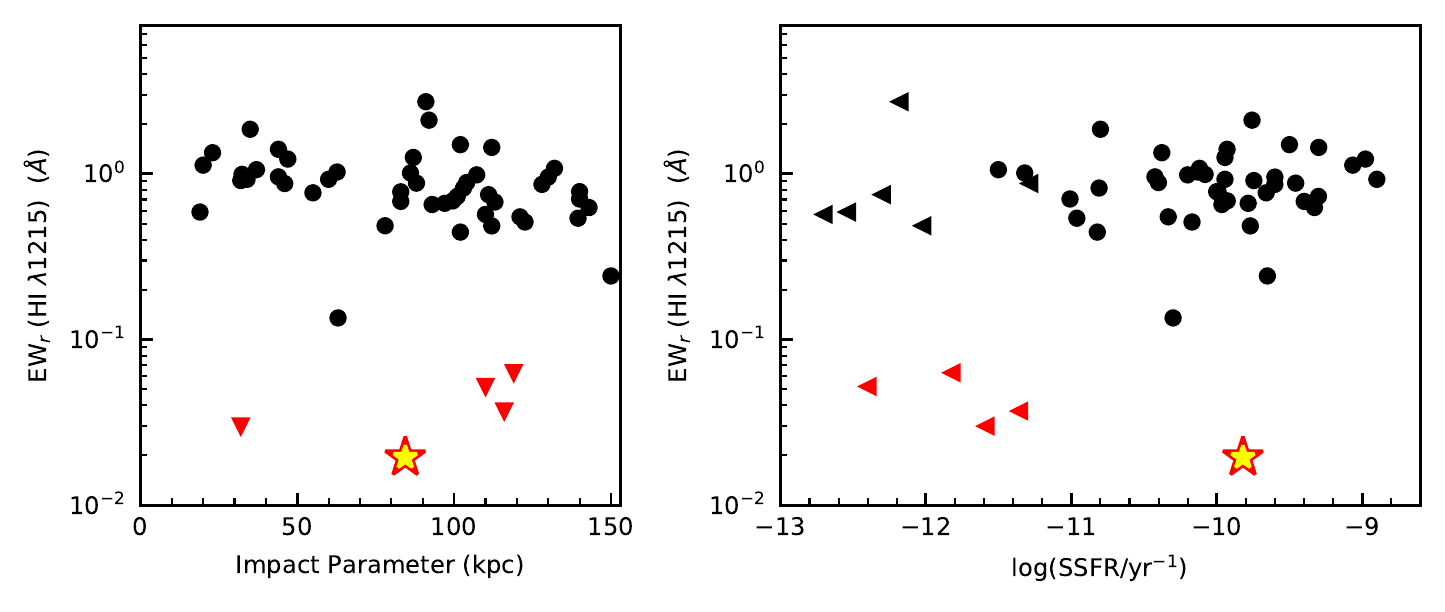}
    \caption{(Left) Subset of galaxies from Fig.~\ref{fig:profiles} that have published SFRs within 150~kpc. We explore the systems with detected {\Lya} absorption (black) and with no detectable absorption (red) within 150 kpc, which contains 5 non-absorbing galaxies including G1547 (star). (Right) The specific star formation rate (SSFR) as a function of {\Lya} equivalent width. Note that 10/12 galaxies below SSFR$\sim$10$^{-11}$~yr$^{-1}$ have SSFR limits, with 4/12 not possessing CGM {\Lya} absorption. Contrarily, G1547 is the only galaxy above 10$^{-11}$~yr$^{-1}$ that is not associated with {\Lya} absorption. It is plausible that G1547 is low star-forming, green-valley galaxy that is becoming quenched due to a lack of CGM.}
    \label{fig:nonabs}
\end{figure*}

\section{Discussion}
\label{sec:discussion}
We have presented an interesting galaxy (G1547) that has no detectable CGM along a quasar sightline at a impact parameter of 84.4~kpc and is one of only two galaxies out of 72 galaxies within 100~kpc that does not exhibit {\HI} absorption. G1547 is a highly inclined disk-dominated galaxy that has its major axis pointed towards the quasar sightline. G1547 has a low star-formation rate and seems to be typical of a green valley galaxy as it resides between the blue and red sequences. Here we explore why this galaxy, and a few others, do not exhibit CGM absorption. 

Galaxy environment could play a critical role in determining the CGM properties of galaxies. Fig.~\ref{fig:env} shows a galaxy census within a 3$'$ radius of the quasar sightline as seen in the Pan-STARRS image, which is plotted next to the spectroscopic and SDSS photometric redshifts as a function of distance from the quasar. The data are colour-coded as a function of their apparent $r-$magnitude and galaxies with photometric redshifts consistent with G1547 have dark grey (for faint galaxies with $m_r\geq20$) and black ($m_r\leq20$) error bars, respectively. The physical scale assumes the redshift of G1547, and this may not be representative of galaxies at different redshifts. The quasar redshift and its environment is also noted. There are five galaxies with spectroscopic redshifts consistent with the quasar redshift, and  many additional galaxies having consistent photometric redshifts with the quasar. Also note, some of the photometric redshift errors are large enough that some galaxies are consistent with some combination of the quasar, G1547 and $z=0.09650$ absorbing galaxy. Therefore, there is only a limited amount of information that we can derive about the environment using photometric redshifts.  

There are five faint dwarf galaxies ($m_r>21.5$) that have photometric redshifts consistent with G1547, with the nearest being at 150~kpc ($\sim60''$, or twice the distance of G1547). Given the magnitude of this object, it would likely contribute little to the total mass of the system and have a virial radius much smaller than G1547. Further note that 4/5 of those dwarf galaxies have photometric redshifts that are also consistent with the absorbing galaxy at $z=0.09650$. There are also four other possible companion galaxies highlighted in Fig.~\ref{fig:env} with their photometric redshifts, $m_r$, and impact parameters (assuming the redshift of G1547) listed. Although there could be a larger-scale association, there is not a more massive galaxy that would dominate this system that is above the mass of G1547. All the other potential members are two times further from the quasar than G1547 and one of those galaxies has a photometric redshift consistent with the absorbing galaxy at $z=0.09650$. Furthermore, the quasar sight-line is well outside the virial radius of all other candidate galaxies, including the dwarfs. In addition, G1547 does not show any morphological disturbances that would indicate possible interactions with other potential group members. Though it is possible that G1547 is located in a loose group, its candidate member galaxies are not more massive and reside far enough away such that we should be able to treat G1547 as an isolated galaxy with respect to the quasar sightline.

Examining the high ionisation phase of the CGM traced by {\OVI}, \citet{pointon17} demonstrated that group galaxies tended to have weaker and smaller velocity width absorption relative to isolated galaxies \citep[also see][]{stocke13,stocke14}. 
This could be attributed to the N({\OVI})--halo mass relation, whereby the most massive systems have hot virial temperatures and there is little {\OVI} to be found, since oxygen resides in even higher ionisation phases \citep{oppenheimer16,ng19}. \citet{burchett16} found similar results for {\CIV}, whereby the detection rate is lower for galaxies found in over-densities and/or residing in groups with dark matter halo masses of $M_{\rm h}> 10^{12.5}$~M$_{\odot}$. If G1547 was in a group environment, then this could explain the lack of absorption found here. However, \citet{burchett16} also reported that contrary to the lack of {\CIV} in groups and massive halos, {\HI} is pervasive in the CGM without regard to mass or environment. This is more in line with the findings for low ionisation states, like {\MgII}, where absorption is found out to larger impact parameters \citep{chen10a,bordoloi14,nielsen18} and the absorption profiles have larger velocity spreads for group galaxies \citep{nielsen18}. The overall lack of multi-phase absorption for G1547  is consistent with the lack of high ionisation gas found in groups but inconsistent with the abundance of low ionisation absorption also found for group galaxies. Thus, this provides additional support that environment may not be a factor for G1547. 

If environment is not a plausible answer for why G1547 does not exhibit CGM absorption, then there must be additional clues in its properties. In Fig.~\ref{fig:nonabs} we explore the properties of the non-absorbing galaxies and absorbing galaxies below an impact parameter of 150~kpc. These galaxies are the subset of those presented in Fig.~\ref{fig:profiles}, which have published SFRs. The 48 galaxies with associated {\Lya} absorption are shown in black while galaxies with no detectable absorption are shown in red. The right panel shows the specific star formation rate as a function of {\Lya} rest-frame equivalent width.
There are 10/12 quiescent galaxies that have SSFR limits below 10$^{-11}$~yr$^{-1}$ with 4/12 not possessing CGM {\Lya} absorption.  The "bonus" galaxy from \citet{werk12} that was discussed in Section~\ref{sec:results} is one of these quiescent galaxies having a log(SSFR$/$yr$^{-1}$)$<-11.6$. So although it resides only 32~kpc from a quasar sightline, it is likely that it does not have any significant CGM since it is quenched. It is also worth noting the close interacting pair of quiescent galaxies from \citet{johnson14}, that also have no detected {\HI}, and have log(SSFR$/$yr$^{-1}$)$\sim$ $-11.2$ and $-11.7$. These galaxies could have either removed the CGM during their interactions, or simply did not have significant CGM to begin with.  

Contrarily, G1547 is the only galaxy of the 40 with SSFR above 10$^{-11}$~yr$^{-1}$ that does not have {\Lya} absorption. It appears that SSFR is a good indicator for the presence of absorption above SSFR~$>10^{-11}$~yr$^{-1}$, while for SSFR~$<10^{-11}$~yr$^{-1}$ there is 67\% chance (8/12) of detecting absorption in quenched/quiescent galaxies. This is consistent with quiescent galaxies containing low ionisation absorption yet still not forming a significant amount of stars, if any \citep{chen19,gauthier09,huang16,thom12,zahedy19LRG}, due to the cold gas never reaching the galaxy because most of it is destroyed in the hot halo \citep{afruni19}. This is also consistent with findings of \citet{tumlinson11}, where they found that galaxies with log(SSFR$/$yr$^{-1}$)<10$^{-11}$ do not exhibit {\OVI} absorption due to a possible lack of outflows.

The fact that G1547 is likely an isolated galaxy that has a lower mass, below the threshold for a shock-heated CGM \citep[e.g., log($M_{\rm h}/$M$_{\odot}) \sim 12$,][]{birnboim03,keres05, stewart11, vandevoort11}, and has a low SFR, SSFR, lack of AGN or star-formation rate-driven outlfows, and is located in between the star-forming sequence and the red sequence, makes it highly probable that this green valley galaxy could be in the process of quenching due to a lack of CGM. It is even more interesting that, given its geometry, the likelihood of detecting absorption would be increased \citep{bouche12,kcn12,kacprzak19a}, typically via accretion, yet none is found.  Having an above average ISM metallicity for its mass and an observable metallicity gradient provides possible evidence for a lack of CGM accretion. Thus, the lack of CGM in this particular object may be the reason for its transformation from the star-forming sequence to the red sequence. 

It is of course also worth contemplating that this particular quasar sightline happens to be sampling the galaxy halo where no CGM is present. High resolution cosmological simulations show that there is a significant amount of variability in the structure of the CGM and the covering fraction of {\HI} gas around star-forming galaxies at $z=0-2$ \citep{hummels19,peeples19,vandevoort19}. However, this structure variability and patchiness is column density dependent. \citet{vandevoort19} examined a simulated Milky Way-mass galaxy at $z=0-0.3$ using both standard mass refinement methods and 1~kpc spatial refinements within the virial radius. They found that the covering fraction of {\HI} gas was unity for log(N({\HI})/cm$^{-2}$)$<13.5$, regardless of the resolution of the simulations. This appears consistent with the simulations of \citet{hummels19} and \citep{peeples19} for higher redshift galaxies. Therefore it is highly unlikely that typical star-forming galaxies would have a non-unity {\HI } covering fraction for low column density limits. This is consistent with Fig.~\ref{fig:nonabs}, where we find that, except for the green valley galaxy G1547, all star forming galaxies exhibit detectable {\HI} absorption.  The simulations show that it is not plausible for G1547 to be a by-chance hole in the CGM and thus could imply a different CGM for it. Hopefully future simulations will explore the CGM within a range of galaxy types in order to understand how the CGM influences galaxy properties. Regardless, even with these simulations one cannot rule out this highly unlikely, yet simple scenario of the quasar sightline passing through a hole in the CGM.  


It is then also plausible that given the nearly edge-on orientation of this galaxy, maybe the cross-section of accreting gas is very small. \citet{kcn12} showed, by examining the {\MgII} absorption azimuthal angle distribution, that the half opening angle of accreting gas along the major axis could be as large as 20 degrees. However, this was derived for a range of galaxy inclinations, which could have resulted in an overestimate of the actual gas cross-section. Given G1547's orientation with respect to the quasar, the accreting filaments would need to have a half opening angle of less than $9.1\pm0.2$ degrees, or a half thickness of $13.4\pm0.9$~kpc (calculated at the quasar sightline 84.4~kpc away from the galaxy). A half opening angle of less than 9.1 degrees for one filament on one side of the galaxy would imply a total gas accretion covering fraction of less than 10\% on the sky if we assume unity gas covering fraction. This low number is consistent with the $\sim$10\% or less found for accreting gas in cosmological simulations \citep{danovich15,fauchergiguere11,vandevoort+schaye12}. However, this is inconsistent with the large number of galaxies that have {\MgII} absorption roughly along the major axis that is co-rotating with their host galaxies \citep[see review by][]{kacprzak17}, which has been predicted to be a signature of gas accretion \citep{danovich15,stewart11,stewart13,stewart17}. 

Overall, given the rarity {\HI} CGM holes detected around galaxies, and given the specific properties of G1547, we conclude that it is likely undergoing quenching due to an absence of fuel from the CGM. 

We can estimate the timescale for G1547 to transition to a quenched galaxy by estimated its total gas mass. We have applied the inverse Kennicutt--Schmidt law following the methods of \citet{papovich15}, which uses the measured galaxy SFRs and sizes to constrain galaxy gas masses. Using the effective radius for G1547 of $3.48\pm0.1$~kpc and its SFR, we compute a gas mass of log(M/M$_{\odot}$)$=9.22\pm0.08$. Given G1547's current SFR, this results in a quenching timescale of $4\pm1$~Gyr. Quenching timescales measured and modelled for galaxies exhibit a large range between 100~Myr to 10 Gyr \citep{correa19,crossett17,janowiecki20,schawinski14,trayford16}. However, disk galaxies tend to have a longer transition time on the order of 2.5~Gyr compared to 100~Myr for early-type galaxies \citep{schawinski14}.  Thus, G1547 has a quenching timescale consistent with disk galaxies in the green valley.

The large range in quenching time-scales highlights the range of external and internal mechanisms that could drive galaxies into or out of the green valley \citep[e.g.,][]{smethurst15}. Given the lack of CGM around G1547, it is plausible that this galaxy is undergoing starvation by not accreting new material to form stars. If there is no longer gas being accreted onto the galaxy, then this would be further suggestive that G1547 may not transition in and out of the green valley, but rather remain on a smooth transition to becoming quenched once the current gas supply is exhausted.

\section{Conclusions}
\label{sec:conclusion}
We have performed a detailed analysis of a unique galaxy (G1547) that does not exhibit any CGM absorption down to very stringent equivalent width limits. It is the only star-forming galaxy within 100~kpc of quasars that does not even have {\Lya} absorption to $EW_r<0.019$~{\AA} $[$log(N({\HI})/cm$^{-2}$)$<12.6]$. Even more particular is that the galaxy is nearly exactly edge-on ($i=86$ degrees) with its major axis pointed towards the quasar ($\Phi=9$ degrees). We have determined the following:

\begin{itemize}
    \item G1547 is an emission-line galaxy that is mildly star forming (SFR$=1.1\pm0.3$~M$_{\odot}$~yr$^{-1}$) with a log(SSFR$/$)$=-9.8\pm0.1$~. G1547's SFR and SSFR places it between the red and blue sequence for galaxies of similar mass and redshift, within the green valley. It has a solar metallicity ISM, which is above average for its mass, and a negative metallicity gradient. G1547 does not exhibit any recent AGN activity as indicated from the BPT diagram.\\
    
    \item G1547 does not exhibit any signatures of outflows as determined via the low star formation surface density ($\Sigma_{\rm
    SFR}$=$0.06\pm0.02$~M$_{\odot}$~yr$^{-1}$~kpc$^{-2}$), the stellar-like {\MgIb} and {\NaID} line ratios, and the low velocity offsets between nebular emission lines, and the {\NaID} stellar+ISM line ($\Delta v =-32\pm19$~{\kms}). Although we comment that constraining the non-presence of stellar driven outflows is more difficult for an edge-on galaxy.\\
    
    \item G1547 does not have any nearby companions based on photometric redshifts. The nearest galaxy of similar magnitude is 2.5 times further from the quasar, with three other similarly bright galaxies at even further distances with uncertain redshifts. Along this quasar sightline, we have concluded that G1547 is relatively isolated. 
    
\end{itemize}

From the properties of G1547, and along with the other absorbing and non-absorbing {\HI} galaxies within 150~kpc of quasar sightlines, we conclude:
\begin{itemize}
    \item High SSFR is a key indicator as to whether or not galaxies exhibit {\HI} absorption. Out of 40 galaxies with SSFR$\geq10^{-11}$~yr$^{-1}$, G1547 is the only galaxy for which no {\HI} absorption is detected (to a stringent upper limit that is a factor of 10 below the weakest detection).  \\
    
    \item On the other hand, when galaxies have a SSFR<10$^{-11}$~yr$^{-1}$, 
     there is 67\% chance of detecting absorption in quenched/quiescent galaxies. This is consistent with the findings that some quiescent galaxies do contain low ionisation absorption yet still are not forming a significant amount of stars \citep{chen19,gauthier09,huang16,thom12,zahedy19LRG}.
\end{itemize}

Our results showing that the SSFR is correlated with the presence of {\HI} is consistent with \citet{tumlinson11} who found that galaxies with SSFR<10$^{-11}$~yr$^{-1}$ do not exhibit {\OVI} absorption. It is still debated as to whether the  \citet{tumlinson11} results are SSFR-driven or halo mass-driven due to the quiescent galaxies in their sample having much higher masses than the star-forming galaxies. Here we have attempted to control for that using a smaller mass range of $9\leq$log(M$_*/$M$_{\odot}$)$\leq$11.

G1547 is the only galaxy (1/40) with a SSFR$\geq10^{-11}$~yr$^{-1}$ that does not contain {\HI} absorption.  Even though cosmological simulations suggest that holes in the CGM do not exist below a column density limit of log(N({\HI})$/$cm$^{-2}$)$<12.6$, we are unable to rule out this possibility. Admittedly this is only one quasar sight-line through a galaxy and we could never conclusively prove or disprove the presence or lack of CGM around the entire galaxy. However, combined with all of G1547's properties and the lack of CGM, we conclude that G1547 is likely undergoing a quenching due to a lack of CGM and is transiting from a star-forming galaxy to a quiescent galaxy, given its location in the green valley. Using all the observable properties together combined with those from the CGM shows the potential power in the ability to further understand how galaxies evolve. Future studies comparing differences in the CGM properties between different galaxy types, including green valley galaxies, will enable us to determine if the CGM plays a critical role in transforming galaxies.  Mapping the CGM or obtaining 21-cm maps would be most helpful in providing a complete picture of how this galaxy is evolving.

\section*{Acknowledgements}
We thank Joe Burchett for discussions and for providing data used in Fig.~\ref{fig:profiles}.
G.G.K. and N.M.N.~acknowledge the support of the Australian
Research Council through {\it Discovery Project} grant
DP170103470. Parts of this research were supported by the Australian
Research Council Centre of Excellence for All Sky Astrophysics in 3
Dimensions (ASTRO 3D), through project number CE170100013. C.W.C. and J.C.C. were
supported by the National Science Foundation through grant NSF
AST-1517816 and by NASA through HST grant GO-13398 from the Space
Telescope Science Institute, which is operated by the Association of
Universities for Research in Astronomy, Inc., under NASA contract
NAS5-26555. T.N acknowledges support from the Australian Research Council Laureate Fellowship FL180100060. S.M is supported by the Humboldt Foundation (Germany) through the Experienced Researchers Fellowship. This research was supported by the Munich Institute for Astro- and Particle Physics (MIAPP) which is funded by the Deutsche Forschungsgemeinschaft (DFG, German Research Foundation) under Germany´s Excellence Strategy -- EXC-2094 -- 390783311. Some of the data presented herein were obtained at the W. M. Keck Observatory, which is operated as a scientific partnership among the California Institute of Technology, the University of California and the National Aeronautics and Space Administration. The Observatory was made possible by the generous financial support of the W. M. Keck Foundation. ESI observations were supported by Swinburne Keck programs 2016A\_W056E, 2015B\_W018E, 2014A\_W178E and 2014B\_W018E. The authors wish to recognise and acknowledge the very significant cultural role and reverence that the summit of Maunakea has always had within the indigenous Hawaiian community. We are most fortunate to have the opportunity to conduct observations from this mountain.

\section*{Data Availability}
The data underlying this paper will be shared on reasonable request to the corresponding author.


\bibliographystyle{mnras}
\bibliography{mnras_template} 







\bsp	
\label{lastpage}
\end{document}